\documentclass{article}

\usepackage{graphicx,color}
\usepackage{amssymb}
\usepackage{amsmath}
\usepackage{textcomp} 
\usepackage{xspace}
\usepackage{focus} 
\usepackage{bbding} 
\usepackage{pifont} 
\usepackage{stmaryrd} 
\usepackage{mdwlist} 
\usepackage{url}
\usepackage{html}

\newcommand{\dto}{\Rightarrow}

\newcommand{\isabh}{Isabelle/HOL\xspace}

\author{Maria Spichkova}
\title{Refinement-Based Specification:\\ Requirements and Architecture}
\begin{document}
\date{\vspace{-5ex}}
\maketitle

\begin{abstract}
This paper presents the methodology for the system requirements and architecture w.r.t. their de\-com\-position and refinement.
It also introduces ideas of  refinement layers and of  refinement-based verification.  
\end{abstract}

\section{Motivation}
\label{intro}

The correctness of a system according to a given specification is essential,
especially for safety-critical applications. 
A formal specification is in general more precise than a natural language one, but 
a formal specification can as well contain mistakes or disagree with requirements: 
it is not enough to have detached formal specifications, 
we also need to validate and to verify them to be sure that 
the specification conforms to its requirements. 

In this paper we focus on the formal specification phase: on requirements specification and on the developing of a logical system architecture and on the corresponding system decomposition. 
There is a large number of approaches to the decomposition methodologies  (see, e.g., \cite{Hofmann_approachesto,PR99,2006-01-1222,TUM-I0818}). 
The main difference and the main contribution of our methodology is that it  was developed for such a system architecture, 
where we have already specified systems or components properties in a formal way  and need to decompose this whole properties collection to a number of subcomponents to get readable and manageable specifications. 
Thus, the presented methodology allows us to decompose system or component architecture exactly on this point where we see that the component specification becomes too large and too complex. 
In many cases the real complexity of a component and, consequently, of its formal specification is realized only during the specification process, when we comes from semiformal 
(or, even harder, from informal) general description to a formal one -- only by collecting and combining all the component properties together for the first time we also get the feeling of the component complexity for the first time. 
Moreover, during this step a number of component properties can added, in most cases some refinement is necessary.  

In the context of hardware and software systems, 
the definition of (formal) \emph{verification} is 
the act of proving or disproving the correctness of a system with respect to 
a certain formal specification or property, using formal methods of mathematics, but we can also 
 see a verification of a system as 
a special case of validation: 
if the property to prove is presented as an abstract specification, 
it remains to validate the system specification with respect to these abstract specification, 
i.e.\ to show that the \emph{refinement relation} holds.
We call this view \emph{refinement-based verification}, and 
present according to these ideas an introduction to  specification groups  and  refinement layers, as well as how these ideas of can be used to optimize the verification process, and which influences it has on the specification process. 

The  feasibility  of this approach was proven on a number of case studies, the most interesting of them, Cruise Control System specification belonging to the Robert Bosch GmbH case study, can be seen at~\cite{spichkova_tb_decomp}: this system has 
75 components (64 atomic components)  and and yields approx. 17 KLOC 
of generated code and 38 KLOC of generated Isabelle/HOL theories, respectively. 

\section{Architecture: Decomposition + Refinement}
\label{sec:methodlology}

Let assume a formal  specification of some component, which covers a large number of its properties, s.t. 
most of which have strong correlation, and 
let this component describes among others the system states and transitions between them, s.t. the resulting representation must correspond to a state transition diagram.  
If we specify this component as a single, non-composite, specification we get a set of formulas that is not really understandable. 
Trying to built a state transition diagram for the whole component, we will get a large automat with spaghetti-transitions between them -- 
this representation will be useless and not manageable. Moreover, the later representation will be not fit the model checker restrictions. 
Therefore, we have a challenge to decompose it  in a number of subcomponents to get some (more) readable specification. 
A simple, intuitive and informal, way to decompose a component is not suitable here. 
In this case we need to have some rules to decompose the component according to the kinds of its logical properties. 
Very important point here is to determine, whether the strong/weak causality property be preserved. 

We start the decomposition to observe the properties that correspond to the different kinds of automats: Mealy and Moore. 
By definition, any state machine can be either a Mealy automat, where the output depends both on the current input and state,  or a Moore automat, where the output depends only from the current state. 
Generally, having a specification represented by a number of formulas, we can divide these formulas into two parts: formulas, which correspond to the definition of a Mealy automat, and formulas, which correspond to the definition of the Moore automat. 
Thus, having a component $CComp$ describing large state machine, we can decompose it into two components by this criteria.

As the next step we propose to use a decomposition schema for all local variables   
that have complicated computation specification: they are moved (together with the according specification parts that describe their computation) via decomposition  from a component $C$ to some extra component $CLoc$.  
Applying the decomposition schema we get two specifications, $C'$ and $CLoc$, which composition results the specification $C$.
After that we propose to use a decomposition schema for  all output streams %
and corresponding 
formulas describing them (depending only on the component state, local variables and some inputs)
that are moved via decomposition  from a component $C'$ to some extra component $COut$. 

The formal decomposition schemas as well as the corresponding case study on Cruise Control System (see technical report \cite{spichkova_tb_decomp}) do not be presented here for lack of space.

\section{Refinement-Based Specification}
\label{section:ideas_refine}

Let a general specification $S_0$ of a system corresponds to the formalization of system requirements. 
In order to show that a concrete specification $S_n$ 
that we get after $n$ refinement steps fulfills the system requirements, we only need to show 
that the specification $S_n$ is a refinement~\cite{focus,broy_refinement2}
 of the specification $S_0$. 
In this context, it is an important point what exactly a developer
means by ``refinement'' on each refinement step (a behavioral refinement, 
an interface refinement, or a conditional refinement, changing time granularity etc.) 
and which specification semantics is used. 
We can see any proof about a system as the proof 
that a more concrete system specification is a refinement of a more abstract one: 
if the property to prove is presented as an abstract specification, 
it remains to validate the system specification with respect to these abstract specification, 
i.e.\ to show that the refinement relation holds (for details see \cite{ms_refine2008}).

\begin{figure}[ht]
\begin{center}
\includegraphics[scale=0.45]{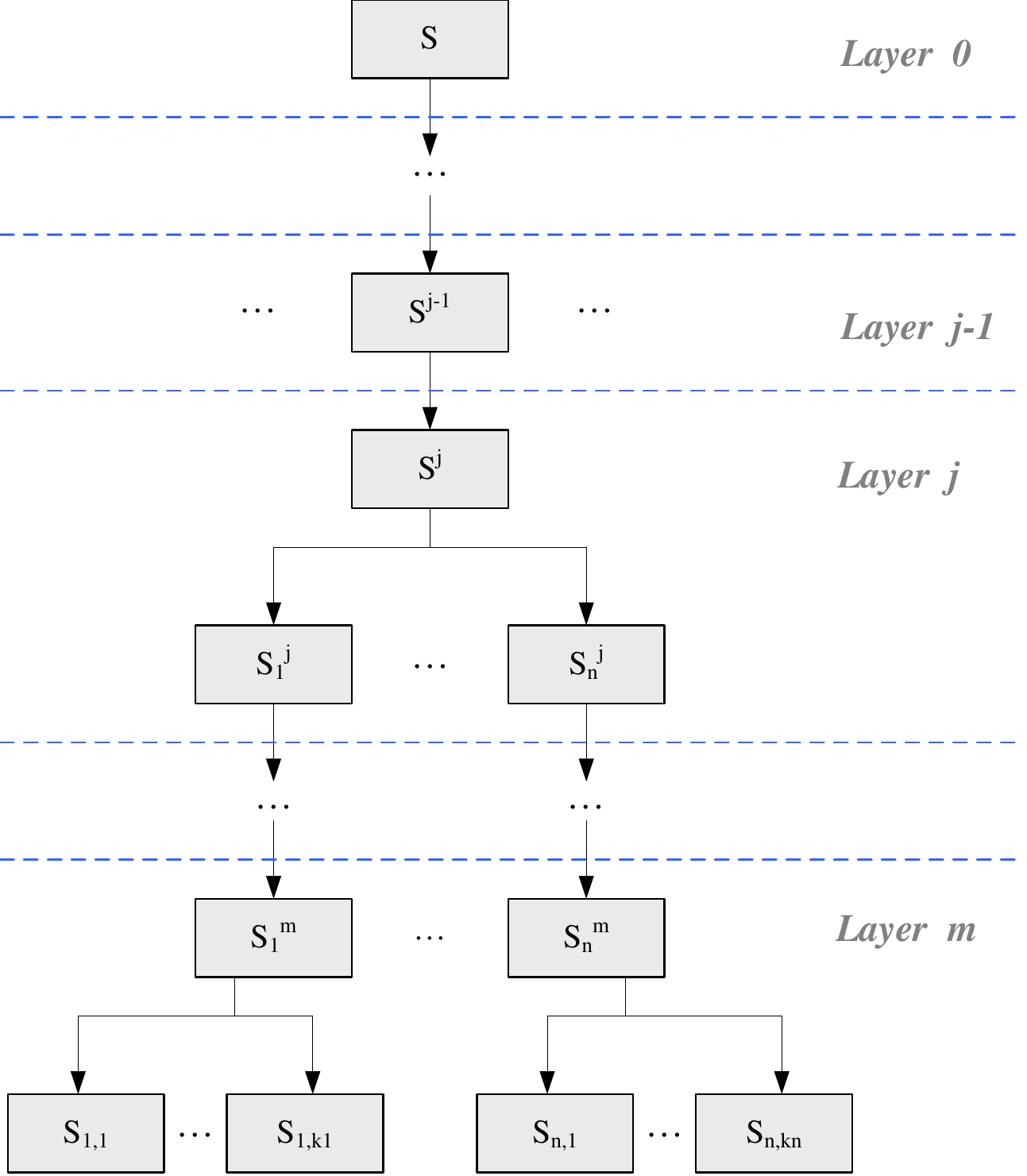}
\end{center}
\caption{Refinement Layers of a Specification Group $S$}
\label{fig:system_layers}
\end{figure}

Fig. \ref{fig:system_layers} represents the hierarchy in a \emph{specification group} $S$ 
in general. The number $N$ of all specification in the group is larger or equal  the number $m$ of refinement layers:
the specification $S^1$ is just a refinement specification of $S$, where $S^j$ is a composition 
		of specifications $S^j_1, \dots, S^j_n$ (where for the specifications 
		$S^j_1, \dots, S^j_n$ the refinement layer $j$ is the most abstract one) 
		that builds a refinement of $S^{j-1}$.   

\begin{figure}[ht]
\begin{center}
\includegraphics[scale=0.5]{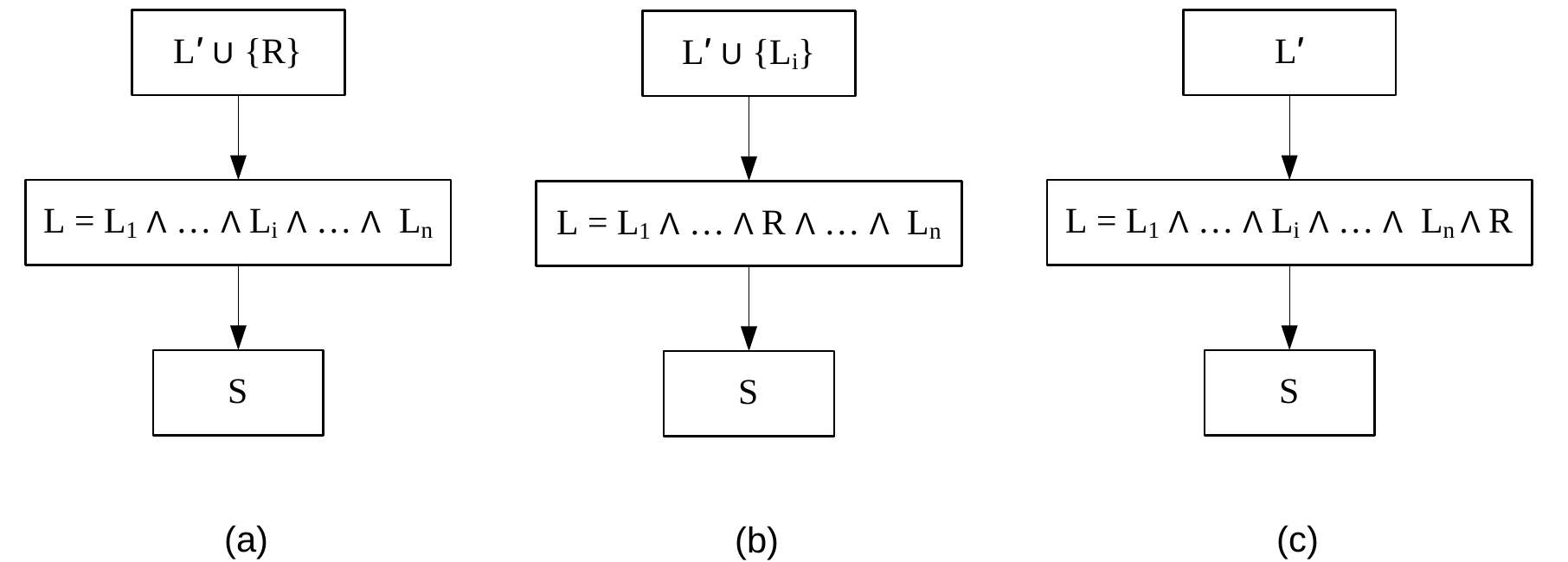}
\end{center}
\caption{Adding new requirement to the list of requirements of the specification}
\label{fig:req_hierarchy}
\end{figure}

\noindent
Assuming a system $S$ with corresponding list of requirements $L = [L_1, \dots, L_n]$:\\
$\semanticsm{S} \Rightarrow \semanticsm{L}$
where 
$
\semanticsm{L} = \semanticsm{L_1} \wedge \dots \wedge \semanticsm{L_n}
$. 
For any new requirement $R$ on the system $S$ that we need to add to the list 
of its requirements $L$, $L \cup \{R\}$ 
(assuming $R$ does not belong to the list of requirements)
we can have the following cases that are intuitively clear.  
	\begin{itemize}
	\item[(1)]
	The system $S$	has some requirement $L_i$ that is less abstract than $R$:\\ 
	$R \notin L ~\wedge~ \exists L_i \in L:  L_i \dto R$.
	\\
  We add $R$ to the next level of abstraction $L'$ 
  (to the list with more abstract requirements, $\semanticsm{L} \Rightarrow \semanticsm{L'}$)  
  using the same schema: $L' \cup \{R\}$, see Fig.~\ref{fig:req_hierarchy}~(a).	
	\item[(2)]
	The list of requirements of the system $S$ has a requirement that 
	is more abstract than $R$:~
	$R \notin L ~\wedge~ \exists L_i \in L: R \dto L_i$.
	\\
	We replace the requirement $L_i$ in $L$ by $R$, 
	$L_i$ will be added to the next level of abstraction $L'$, see Fig.~\ref{fig:req_hierarchy}~(b). 	
	If $S$ does not fulfill $R$, 
	then $S$ must be changed according to the new list of requirements.
	\item[(3)]
	The system $S$ has no requirements that are in some relation (more/less abstract) to $R$ 
	($R$ opens some new ``dimension'' of $S$):\\ 
	$R \notin L ~\wedge~ \forall L_i \in L: \neg (L_i \dto R) \wedge \neg (R \dto L_i)$. 
	\end{itemize}

\noindent
A list of requirements can also be represented by a formal specification. 
Thus, we allude the refinement layers (see Fig.~\ref{fig:system_layers}). 
If the requirement specification can be extended, we always have a choice: 
 either we extend the specification itself and don't make any changes of the refinement layers
 or we don't make any changes of the original specification, 
	but add some new refinement layer with the extended version of the specification.

\section{FOCUS on Isabelle}
\label{section:focus_main}

The main ideas, presented in the paper, are language independent, but for the better readability and for better understanding of this ideas we shoe them ob the base of formal specifications presented in the \Focus~\cite{focus}, a framework for formal specifications and development of interactive systems.\footnote{
See \url{http://focus.in.tum.de}.
}
We can also see this methodology as an extension of the approach ``\Focus on Isabelle''~\cite{spichkova} -- it is integrated into a seamless development process\footnote{%
See Verisoft-XT project, \url{http://www.verisoftxt.de}.%
}, 
which covers both specification and verification, starts  from informal specification and finishes by the corresponding verified C code. 
 Given a system, represented in \Focus, one can verify its properties by translating the specification
to a Higher-Order Logic and subsequently using the theorem prover Isabelle/HOL or the point of disagreement can be found. 
For a detailed description of Isabelle/HOL see \cite{npw} and \cite{IsabelleManual}.

\Focus is preferred here over other specification frameworks since it 
has an integrated notion of time and modeling techniques for unbounded networks, provides a number of specification techniques for distributed systems and concepts of refinement. 
For example, the B-method~\cite{bbook} is used in many publications 
on fault-tolerant systems, but 
it has neither graphical representations nor integrated notion of time. 
Moreover, the B-method also is slightly more low-level and more focused on the refinement to code rather than formal specification.  
Formal specifications of real-life systems can become very large and complex, 
and are as a result hard to read and to understand. 
Therefore, it is too complicated to start the specification process in some low-level framework directly. 
To avoid this problem \Focus supports a graphical specification style 
based on tables and diagrams. %

The main point in ``\Focus on Isabelle'' is an alignment on the future proofs 
to make them simpler and appropriate for application not only in theory but also in practice.  
The proofs of some system properties can take considerable (human) time 
since the  \isabh is not fully automated. 
But considering ``\Focus on Isabelle''
we can influence on the complexity of proofs already doing the specification 
of systems and their properties.
Thus, the specification and verification/validation methodologies are treated as 
a single, joined, methodology with the main focus on the specification part. 

In addition, the methodology  helps to perform the next modeling step -- translation to the  case tool representation and deployment: we can schematically translate the \Focus specification to a model in AutoFocus~3~\cite{SchaetzHuber:TR2001}, 
a tool for modeling and analyzing the structure and behavior of distributed, reactive, and timed computer-based systems.\footnote{%
See \url{http://af3.in.tum.de}.
} 
Having such a model we can simulate it, prove its properties using model checking and also using its translation to Isabelle/HOL, as well as we gan generate C code from it.

\section{Conclusions}
This paper presents the methodology for the system requirements and architecture w.r.t. their de\-com\-position and refinement.
The main contribution of our decomposition methodology is that it  was developed for such a system architecture, where we know systems properties and need to decompose the whole properties collection to a number of subcomponent. 
Thus, the presented methodology allows us to decompose system or component architecture exactly on this point where we see that the component specification becomes too large and too complex to work with it.  
In addition, our methodology helps to perform the next modeling step -- translation to the  case tool representation and deployment.

This paper introduces also briefly the ideas of  specification groups  and  refinement layers, as well as how the ideas of 
the refinement-based verification can be used to optimize the verification process, and which influences it has on the specification process. We can also see the presented methodology as an extension of the approach ``\Focus on Isabelle''~\cite{spichkova} -- it is integrated into a seamless development process, which covers both specification and verification, starts  from informal specification and finishes by the corresponding verified C code.%
 
\bibliographystyle{plain}

\begin{thebibliography}{10}

\bibitem{bbook}
J.-R.\ Abrial.
\newblock {\em {The B-book: assigning programs to meanings}}.
\newblock Camb.\,\!Univ.\,\!Press, 1996.

\bibitem{broy_refinement2}
M.~Broy.
\newblock Compositional refinement of interactive systems modelled by
  relations.
\newblock {\em COMPOS'97: Revised Lectures from the International Symposium on
  Compositionality: The Significant Difference}, pages 130--149, 1998.

\bibitem{focus}
M.~Broy and K.~St{\o}len.
\newblock {\em Specification and Development of Interactive Systems: Focus on
  Streams, Interfaces, and Refinement}.
\newblock Springer, 2001.

\bibitem{TUM-I0818}
D.B. da~Cruz and B. Penzenstadler.
\newblock {Designing, Documenting, and Evaluating Software Architecture}.
\newblock Technical Report TUM-I0818, TU M{\"u}nchen, 2008.

\bibitem{Hofmann_approachesto}
C.~Hofmann, E.~Horn, W.~Keller, K.~Renzel, M.~Schmidt, W.~Horn, and B.~Anger.
\newblock Approaches to software architecture.

\bibitem{SchaetzHuber:TR2001}
F.~Huber and B.~Sch\"atz.
\newblock {Integrated Development of Embedded Systems with \textsc{AutoFocus}}.
\newblock Technical Report {TUMI-0701}, {TU M{\"u}nchen}, 2001.

\bibitem{npw}
T.~Nipkow, L.~C. Paulson, and M.~Wenzel.
\newblock {\em {Isabelle/HOL -- A Proof Assistant for Higher-Order Logic}},
  volume 2283 of {\em LNCS}.
\newblock Springer, 2002.

\bibitem{PR99}
J.~Philipps and B.~Rumpe.
\newblock {Refinement of Pipe-and-Filter Architectures}.
\newblock In J.~M. Wing, J.~Woodcock, and J.~Davies, editors, {\em World
  Congress on Formal Methods (FM'99)}, number LNCS 1708, pages 96 -- 115.
  Springer, 1999.

\bibitem{spichkova}
M.\ Spichkova.
\newblock {\em {Specification and Seamless Verification of Embedded Real-Time
  Systems: FOCUS on Isabelle}}.
\newblock PhD thesis, {TU M{\"u}nchen}, 2007.

\bibitem{ms_refine2008}
M.~Spichkova.
\newblock Refinement-based verification of interactive real-time systems.
\newblock In {\em {REFINE 2008}}. ENTCS, 2008.

\bibitem{spichkova_tb_decomp}
M.~Spichkova.
\newblock {Architecture: Methodology of Decomposition}.
\newblock Technical Report {TUM-I1018}, {TU M{\"u}nchen}, 2010.

\bibitem{IsabelleManual}
M.~Wenzel.
\newblock {\em The Isabelle/Isar Reference Manual}.
\newblock TU M\"unchen, 2004.

\bibitem{2006-01-1222}
D.~Wild, A.~Fleischmann, J.~Hartmann, C.~Pfaller, M.~Rappl, and S.~Rittmann.
\newblock {An Architecture-Centric Approach towards the Construction of
  Dependable Automotive Software}.
\newblock In {\em Proceedings of the SAE 2006 World Congress}, 2006.

\end{thebibliography}

\end{document}